# Nitrogen-monovacancy ($V_N$) Hexagonal Boron Nitride 2D Monolayer Material as an Efficient Electrocatalyst for $CO_2$ Reduction Reaction

## Lokesh Yadav[1] and Srimanta Pakhira[1,2*]

[1] Theoretical Condensed Matter Physics and Advanced Computational Materials Science Laboratory, Department of Physics, Indian Institute of Technology Indore (IIT Indore), Simrol, Khandwa Road, Indore-453552, Madhya Pradesh, India.

[2] Theoretical Condensed Matter Physics and Advanced Computational Materials Science Laboratory, Centre for Advanced Electronics (CAE), Indian Institute of Technology Indore, Simrol, Khandwa Road, Indore-453552, Madhya Pradesh, India.

*Corresponding author: spakhira@iiti.ac.in (or) spakhirafsu@gmail.com

## Abstract

The conversion of waste carbon dioxide ($CO_2$) gas into valuable products and fuels through an electrocatalytic $CO_2$ reduction reaction ($CO_2RR$) is a promising approach. The sluggish kinetics of the $CO_2RR$ require the development of novel strategies for electrocatalyst design. Two-dimensional (2D) materials emerge as promising candidates for $CO_2RR$ due to their distinctive electronic and structural properties. This study follows the first principles based DFT-D method to examine the electrocatalytic competences of the defective two-dimensional boron nitride monolayer (d-BN) material towards $CO_2RR$. Introducing a particular defect with nitrogen vacancies in the 2D single layer pristine hexagonal boron nitride ($V_N$_d-BN) can efficiently activate the $CO_2$ molecules for hydrogenation by reducing the electronic band gap of the pristine hBN from 6.23 eV to 3.0 eV. Therefore, $V_N$_d-BN material can act as a large band gap semiconductor. Our findings demonstrate that the defective regions in the 2D monolayer $V_N$_d-BN serve as active sites (Boron) for both the adsorption and activation of $CO_2$. The subsequent hydrogenation steps occur sequentially once the $CO_2$ molecule is adsorbed on the catalytic surface. Our results indicate that the OCHO* path is the most favorable for $CH_4$ production. Hence, the 2D monolayer $V_N\_d$-BN material holds a great promise as a cost-effective catalyst for $CO_2RR$, and it presents a viable alternative to expensive platinum (Pt) catalysts.

## Introduction

The rapid growth of industrialization leads to technological advancements in our daily lives, but it also raises significant environmental concerns, primarily due to the increasing pollution levels. The primary source of this increasing pollution level is the combustion of fossil fuels, which alternatively intensifies the $CO_2$ emissions into our environment and subsequent climate change.[1–3] Various strategies have been developed to control the unnecessary production of $CO_2$ emissions, such as minimizing fossil fuel usage by adopting alternate renewable energy sources. An electrochemical $CO_2RR$ is considered one of the most promising techniques for transforming $CO_2$ into valuable products such as methanol ($CH_3OH$), methane ($CH_4$), and formic acid ($HCOOH$) under room temperature and ambient pressure conditions.[4–7] This significant step towards this endeavor for electrochemical conversion of $CO_2$ into valuable products attracts the scientific community for future challenges. Integrating electrochemical $CO_2RR$ into renewable energy sources holds the promise of establishing a carbon-neutral energy cycle.

Over the past decades, researchers have extensively explored photochemical, thermochemical, and electrochemical methods for $CO_2$ reduction.[8–12] Thermochemical conversion, which relies on high temperatures, pressures, and equivalent amounts of hydrogen as a reducing agent, makes it impractical for large-scale applications due to energy constraints.[7,12] In the photochemical processes, a few catalysts have shown some activity in $CO_2RR$, but their selectivity and production rates are prohibitively low from an economic perspective.[8,10] In contrast, electrochemical $CO_2RR$ holds various advantages.[11,13–15] One of the advantages is that it can perform under ambient conditions, and we can precisely control the rate of reaction by adjusting the external bias. Also, the products are generated at the different electrodes, naturally facilitating the separation of products using the individual reaction chambers.[16] This feature reduces the cost of post-reaction separation processes, making electrochemical $CO_2RR$ a sustainable option. However, several challenges remain in the field of electrochemical $CO_2RR$, including high overpotential, low selectivity, and low yields resulting from the thermodynamic stability of $CO_2$. Furthermore, the use of $CO_2$ electrocatalysis is limited due to the excessive costs and inaccessibility of novel metal catalysts such as Pt, Au, and Pd.[17] Many efforts have been made to develop novel, feasible, and efficient electrocatalytic materials that can accomplish high rates of $CO_2RR$ with low overpotential as one of the solutions to these issues. Numerous 2D nanosheets can show promise for the reduction of $CO_2$, depending on the preferred reaction products such as CO, HCOOH, $CH_3OH$, and $CH_4$.

The research on 2D nanosheets following the discovery of graphene has expanded significantly over the past decade, and various 2D materials have been identified, such as covalent organic frameworks, chalcogenides, oxides, nitrides, halides, carbides, hydrides, hydroxides, phosphonates, phosphates, metals, and elements from groups IV and V.[18] These 2D materials can exhibit insulating, semiconducting, and metallic properties. Notably, when multi-layered materials are reduced into a monolayer, they exhibit enhanced and novel electronic characteristics due to quantum confinement. Additionally, 2D materials offer a high specific surface area due to their reduced dimensionality. Different methods, such as doping, intercalation, alloying, and chemical functionalization, help us tune the physicochemical properties of 2D materials.[18] These strategies enhance the opportunities to design nanosheets for catalysis as well as energy applications. Catalysis of the 2D nanosheets has gained considerable attention due to their distinctive structural and electronic properties.[19] These ultrathin nanosheets possess exposed surface atoms on both sides and can escape the lattice to create vacancy-type defects. The presence of vacancy defects in nanosheets can result in the reduction of the coordination number of surface atoms, ultimately boosting catalytic performance. Also, the increment in low-coordinated surface sites enhances the chemisorption of reactants. Controlling vacancy defects can help us to modify the catalytic activities and the corresponding electronic structure. Moreover, the atomic sites present at the low coordinated edges of nanosheets exhibit intriguing catalytic properties.

The remarkable properties of 2D materials (especially hexagonal boron nitride (hBN)) have gained significant attention in various scientific fields. For example, the 2D material hBN has numerous applications, such as biomedical devices, power devices, fuel cells, dielectric tunnelling, stretchable optoelectronic devices, and electrocatalytic water splitting.[20–24] The geometrical structure of hBN closely resembles graphite. A monolayer of hBN can be simulated by substituting all the carbon atoms present in graphene with boron (B) and nitrogen (N) atoms.[25] The alternating B/N atoms are covalently bound into a hexagonal layer, and weak van der Waals forces hold these layers together. Unlike graphite, hBN exhibits a distinct interlayer stacking pattern where B atoms align directly below or above the N atoms in adjacent layers. This configuration indicates the polar nature of B-N bonds due to the distinct electronegativity of B and N atoms, inducing a partial ionic character within these covalent bonds.[26] Due to the higher electronegativity of the N atoms, the electron pairs in $sp^2$-hybridized B–N $\sigma$ bonds are confined to the N atoms. The pristine hBN exhibits exceptional stability, a flat surface, and insulating behavior with a relatively wide bandgap ranging from 3.60 eV to 7.0 eV, depending on the use of experimental methods.[27–30] This wide bandgap of the material is a challenge for utilizing hBN-based materials as efficient electrocatalysts for the $CO_2RR$.

Several strategies, such as metal doping, functional group adsorption, and hydrogenation, have been explored to enable effective electronic communication with the hBN surface.[31,32] Unfortunately, these approaches often lead to structural deformations in the pristine 2D planar geometry of hBN due to a mismatch between the size of the metal atom and the B/N atom. Consequently, these modifications may compromise the desired properties of hBN. The hBN material exhibits considerable potential due to its robust chemical stability, making it a noteworthy material explored across various domains, including nitrogen reduction reaction (NRR), $CO_2$ reduction reaction, and CO oxidation.[33–38] Chen et al. explored a number of metal-loaded BN, revealing that $MoN_3$/BN is a standout performer in NRR by strategically replacing a B atom with Mo on hBN.[38] Ajayan et al. utilized porous hBN as a substrate, employing a one-step vacuum filtration process to successfully fabricate a monoatomic Ru/hBN catalyst for the $CO_2RR$.[35] Furthermore, the defects induced in hBN significantly reduce the band gap, enhancing electrical conductivity to process electrochemical reactions. In other words, it has been observed that the 2D monolayer hBN, in the presence of vacancy defects, gives rise to semiconducting essence.

The 2D pristine hBN nanomaterial cannot be used effectively as an electrocatalyst for the $CO_2RR$ due to its wide electronic band gap (Eg), showing its electrical insulating behavior. Due to the lack of a sufficient number of active catalytic sites, it hinders the practical use of catalytic reactions. A recent experiments performed by Yu Lei et al. demonstrated various vacancies or defects in the pristine 2D single layer hBN material (noted by d-BN), and they found that the presence of different types of vacancies within d-BN induces substantial changes in its band structure and causes a shift in the Fermi energy ($E_F$) level.[39] Furthermore, spin density calculations and electron spin resonance (ESR) experiments provided a theoretical and experimental confirmation of generating localized free radicals in these vacancies.[39] In another study, Katerina L. et al. experimentally showed that the 2D monolayer d-BN with nitrogen defects can activate the $CO_2$ molecule for hydrogenation, and the d-BN sheet can potentially serve as an active electrocatalyst for $CO_2RR$.[40] They found that the d-BN with nitrogen defect catalyzes formic acid and methanol formation at different temperatures.[40] To support and explain the experimental observation, , we have computationally performed our calculations under standard conditions (T = 298.15 K, p = 1 bar, U = 0 V, and pH = 0).

We have theoretically designed a 3x3 supercell of a 2D pristine hBN sheet and introduced a specific defect in this sheet with a single nitrogen-vacancy ($V_N$), forming a nitrogen defective 2D monolayer boron nitride material noted by $V_N\_d$-BN. After that, we studied the electronic and geometrical properties of this material by employing first-principles based dispersion-corrected density functional theory method (DFT-D). This $V_N$ vacancy defect has rendered $V_N\_d$-BN material chemically active. The formation of 2D monolayer $V_N\_d$-BN material from the 2D single layer pristine hBN is depicted in Fig. 1. Using the same DFT-D approach, we calculated the electronic properties such as the electronic band structure, total density of states (DOS), electronic band gap ($E_g$), and position of the Fermi energy level ($E_F$) to study the properties of the $V_N\_d$-BN material with their $CO_2RR$ mechanism precisely. We also performed electron spin density calculations to locate the presence of unpaired electrons in the system. In our previous work, we found that the calculated electronic band gap of the 2D monolayer pristine hBN was about 6.23 eV, but in our current study, we have found that it is reduced to 3.0 eV after introducing the $V_N$ vacancy defect to the pristine hBN which is consistent with the previous study.[41] This reduction in the band gap indicates a novel strategy to improve the $CO_2RR$ performance of the $V_N\_d$-BN material which is well harmonized with the experimental observation.[40] We investigated the $CO_2RR$ mechanism on the surface of 2D single layer $V_N\_d$-BN material. Our research thoroughly examined how $CO_2$ is converted into various products on the catalytic surface using a first principles-based DFT-D approach which explain the experimental results obtained by Chagoya et al.[40] We carefully analyzed the adsorption configurations of $CO_2$ and its intermediates during the $CO_2RR$ process by calculating the changes in Gibbs free energy ($\Delta G$). Our calculations outlined the essential stages of hydrogenating $CO_2$ into C1 products, such as CO, HCOOH, $CH_3OH$, and $CH_4$, indicating that the 2D monolayer $V_N\_d$-BN material exhibits higher $CO_2RR$ selectivity towards $CH_4$ formation. The findings of this study confirm that the 2D monolayer $V_N\_d$-BN can efficiently serve as an electrocatalyst for the $CO_2RR$.

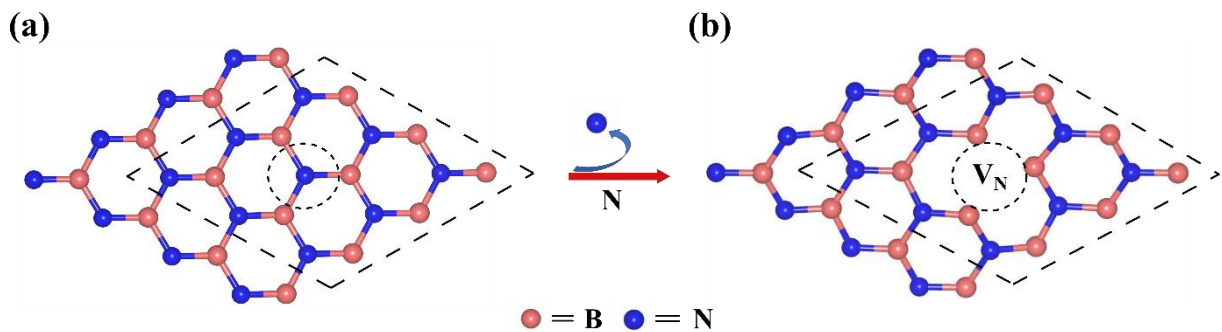

**Fig. 1** Equilibrium structures of (a) the pristine 2D monolayer hBN and (b) the 2D monolayer $V_N$_d-BN materials. (The dotted lines represent the boundary of 3 × 3 supercells.)

## Theoretical Methodology and Computational Details

### (a) Methodology

We investigated the equilibrium structures and properties of the 2D monolayer $V_N$_d-BN material by employing a periodic hybrid dispersion-corrected first principles-based B3LYP-D3 density functional theory (DFT-D) method. which.[42–45] We performed these calculations by using *ab initio* based CRYSTAL17 suite code, which employs Gaussian types of atomic basis sets during the computations. It is more efficient than plane wave-based codes for hybrid DFT-D calculations.[46–51] We used specific Gaussian basis sets with triple-$\zeta$ valence and polarization quality (TZVP) for oxygen (O), hydrogen (H), carbon (C), boron (B), and nitrogen (N) atoms in the present investigation. Due to weak van der Waals (vdW) interactions among the atoms within the layers of $V_N$_d-BN nanosheet, we have also considered the significance of vdW dispersion effects. To address long-range vdW interactions, Grimme's 3rd order (-D3) corrections were incorporated in these computations.[52] The B3LYP-D3 method is helpful in providing reliable and appealing geometries for the monolayer 2D structures because energy and density have the minimum effect of spin contamination in our calculations, where we set the electronic self-consistency scale to $10^{-7}$ atomic units. Two-dimensional vacuum slabs were constructed for the materials to incorporate the electrostatic potential in the computations. We take energy calculations with respect to the vacuum where the height of the vacuum cell is set at 500 Å. We employed a 4 × 4 × 1 Monkhorst–Pack k-mesh grid where we sampled all integrations of the first Brillouin zone.[53] We have set a $10^{-7}$ a.u. threshold for convergence of energy and electron density. We have also used the (VESTA) visualization code for the graphical analysis of all the optimized structures.[54]

**(b) CO₂RR mechanism**

In the CO₂RR process, electrons move toward the cathode through an external circuit, and each electron, together with a proton, reaches toward the $CO_2$ molecule and makes a bond with the carbon or oxygen atom, forming various products. These products can be formic acid (HCOOH), carbon monoxide (CO), methanol ($CH_3OH$), and methane ($CH_4$), based on the paths followed. The primary two-electron ($2e^-$) reduction products in CO₂RR are carbon monoxide (CO) and formic acid (HCOOH), as shown in eqn. (1) and (2). Initially, the hydrogenation of adsorbed $CO_2$ molecule results in the formation of either COOH* or OCHO*, which can further be reduced to CO* or HCOOH*, respectively. As illustrated in the equations below, the intermediates CO* and HCOOH* can then leave the catalyst surface, and one can get CO and HCOOH molecules as final products, respectively, during CO₂RR. Here "*" represents the active site on the surface of the catalyst.

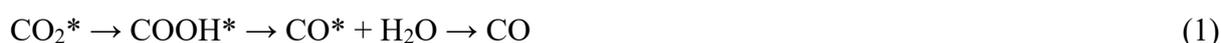
$$CO_2^* \rightarrow COOH^* \rightarrow CO^* + H_2O \rightarrow CO \tag{1}$$

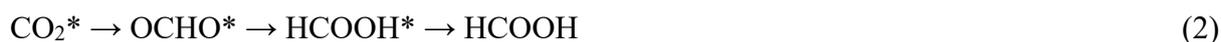
$$CO_2^* \rightarrow OCHO^* \rightarrow HCOOH^* \rightarrow HCOOH \tag{2}$$

The further hydrogenation of adsorbed CO* to COH* or CHO* can lead to the generation of $CH_3OH$ and $CH_4$ molecules as the six-electron or eight-electron reduction products, respectively. To produce a $CH_3OH$ molecule as a final product, eqn. (3) presents the pathway where initially CHO*/COH* reduced to CHOH*, which further reduced to CH₂OH*. Finally, *CH₂OH is reduced to give $CH_3OH$ as the ultimate product.

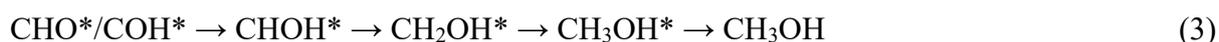
$$CHO^*/COH^* \rightarrow CHOH^* \rightarrow CH_2OH^* \rightarrow CH_3OH^* \rightarrow CH_3OH \tag{3}$$

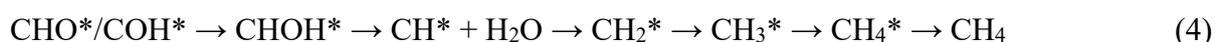
$$CHO^*/COH^* \rightarrow CHOH^* \rightarrow CH^* + H_2O \rightarrow CH_2^* \rightarrow CH_3^* \rightarrow CH_4^* \rightarrow CH_4 \tag{4}$$

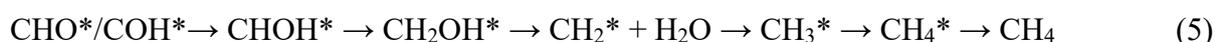
$$CHO^*/COH^* \rightarrow CHOH^* \rightarrow CH_2OH^* \rightarrow CH_2^* + H_2O \rightarrow CH_3^* \rightarrow CH_4^* \rightarrow CH_4 \tag{5}$$

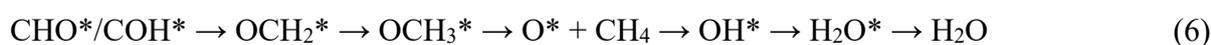
$$CHO^*/COH^* \rightarrow OCH_2^* \rightarrow OCH_3^* \rightarrow O^* + CH_4 \rightarrow OH^* \rightarrow H_2O^* \rightarrow H_2O \tag{6}$$

The formation of a $CH_4$ molecule as a final product can occur through three possible routes, as shown in eqn. (4), (5), and (6). Eqn. (4) involves the initial reduction of CHO* or COH* to CHOH*, followed by hydrogenation to CH* intermediate and $H_2O$ molecule. Subsequently, CH* undergoes further reduction to $CH_4$ through a series of hydrogenation steps. However, it is important to emphasize that the hydrogenation of CHOH* to the formation of CH* intermediate usually presents a high energy barrier compared to the formation of CH₂OH* and OCH₃* intermediates, which indicates that the path in eqn. (4) is less favorable than that mentioned in eqn. (5) and (6) when the CO$_2$RR takes place on the catalytic surface. In eqn. (5), the intermediate CHO* or COH* is initially reduced to CHOH*, which further hydrogenates to CH$_2$OH* and then generates a CH$_2$* intermediate along with a water molecule. Next, the CH$_2$* intermediate undergoes hydrogenation to CH$_3$*, producing CH$_4$ as a final product which leaves from the catalyst surface. On the other hand, eqn. (6) presents an alternative path where CHO* is reduced to OCH$_2$* and further undergoes hydrogenation to form an OCH$_3$* intermediate. The OCH$_3$* intermediate is further reduced into a CH$_4$ molecule in the next step, leaving a single oxygen atom (O*) on the catalytic surface. Eventually, the remaining O* on the catalytic surface is reduced to OH*, and at last, during further hydrogenation, an H$_2$O molecule is produced as a final product.

**(c) Thermodynamic analysis and energy calculations**

The DFT-D approach was consistently applied throughout the calculations to accurately account for the changes in Gibbs free energy ($\Delta$G) during the CO$_2$RR process. The catalytic performance of the V$_N$_d-BN material in the CO$_2$RR was studied by computing the change in Gibbs free energy ($\Delta$G) for each intermediate species involved in the subject reaction. We use the *Computational Hydrogen Electrode* (CHE) model to determine the values of $\Delta$G of each intermediate species during the CO$_2$RR process. We performed all of these calculations under standard conditions (T = 298.15 K, p = 1 bar, U = 0 V, and pH = 0). We have followed the method proposed by Nørskov et al., which demonstrated that the chemical potential of a sum of both electron and proton (H$^+$ + e$^-$) can be correlated with the chemical potential of ½ H$_2$ molecule in the gaseous state utilizing the standard hydrogen electrode.[55] Here, we use this correlation under standard conditions to calculate the energy change for the hydrogenation in each reaction step of CO$_2$RR. We calculated the change in adsorption energies ($\Delta$E) by determining the energy difference between the model with the adsorbed species [E$_{slab + adsorbate}$], the catalytic model V$_N$_d-BN [E$_{adsorbent}$], and the adsorbate itself [E$_{adsorbate}$], as given below:

$$\Delta E = E_{slab + adsorbate} - (E_{adsorbent} + E_{adsorbate})$$

The negative value of change in adsorption energy represents the stability of intermediates on the catalytic surface, indicating that the adsorbate is energetically bound to the surface of V$_N$_d-BN material. Hence, this negative value of change in adsorption energy is favorable for elementary reactions on the surface of V$_N$_d-BN. Additionally, we computed the changes in Gibbs free energy ($\Delta$G) for each reaction step of the CO$_2$RR occurring on the surface of V$_N$_d-BN material using the following equation:

$$\Delta G = \Delta E + \Delta E_{ZPE} - T\Delta S + \Delta G_{pH}$$

Here, $\Delta E$ represents the change in adsorption energy, $\Delta E_{ZPE}$ represents the zero-point energy, $\Delta S$ is the entropy correction (differences between the gas phase and the adsorbed state), and T represents the temperature of the system (we considered here room temperate (T = 298.15 K) for this work). The term $\Delta G_{pH}$ represents the Gibbs free energy change resulting from changes in $H^+$ concentration in an acidic medium, and for this study, we set a zero value to the term $\Delta G_{pH}$.

## Results and discussions

To investigate the $CO_2RR$ mechanism, we computationally designed a 3x3 supercell of a 2D monolayer pristine hBN sheet containing totally number of nine B atoms and nine N atoms. The, we introduced a single nitrogen-vacancy defect ($V_N$) in this 2D monolayer pristine hBN sheet to form a 2D monolayer $V_N$_d-BN material. The equilibrium structures of both the 2D pristine hBN and $V_N$_d-BN materials are shown in Fig. 1a and 1b, respectively. In our recent previous work, we already discussed both the structural and electronic properties of the pristine 2D monolayer hBN material in detail.[41] Recent studies have demonstrated that introducing vacancy defects modifies the physiochemical and electronic structure of materials, improving electrocatalytic performance through defect engineering.[39-41] The catalytic activity of the d-BN depends on the distribution and position of defects within the structure. It is already experimentally and theoretically confirmed by ESR and spin density calculations that these defects generate localized free radicals.[39] We studied the structural and electronic properties of the 2D monolayer $V_N$_d-BN material to analyze the electrocatalytic performance towards the $CO_2RR$. The electronic properties, i.e., electronic band structure and total density of states of the 2D monolayer $V_N$_d-BN material are shown in Fig. 7a. Our current DFT-D analysis indicates the values of lattice constants calculated to be a = 7.285 Å, b = 7.438 Å along with the interfacial angles $\alpha = \beta = 90°$ and $\gamma = 119.32°$ with the P1 symmetry. The $V_N$_d-BN sheet exhibits an electronic band gap of 3.0 eV, which results in a large band gap semiconductor material. The reduced electronic bandgap of $V_N$_d-BN material raises the electron concentration and improves the conductivity to perform $CO_2RR$. It may be helpful to facilitate the electron transfer toward reactants in $CO_2RR$ processes. It offers a novel strategy for effectively utilizing the $V_N$_d-BN material to perform the $CO_2RR$. Consequently, we expect this 2D monolayer $V_N$_d-BN material to demonstrate an improved electrocatalytic activity for $CO_2RR$.

The vibrational stability of a system is a crucial aspect determining its overall stability, especially in relaxed configurations. Raman spectroscopy provides valuable insight into the vibrational modes of molecules of materials, offering a detailed picture of their structural dynamics. Under the DFT-D approaches, we use the B3LYP-D3 method implemented in the CRYSTAL17 suite code to simulate the Raman spectrum, which helps us to study the stability of a system, Raman active frequencies, and amplitudes of vibrational modes under various conditions. For these calculations, we have used a set of 4 x 4 x 1 Monkhorst–Pack k-point grids. We have performed the harmonic vibrational analysis with thermochemistry at room temperature (T = 298.15 K) to check the thermal stability of the system. This harmonic vibrational analysis of the equilibrium structure of both the 2D monolayer pristine hBN and $V_N$_d-BN materials help to study Raman spectra. Fig. 2 shows the simulated Raman spectrum for both (a) hBN and (b) $V_N$_d-BN materials. In the simulated Raman spectrum of the pristine 2D single layer hBN, two distinct peaks are observed at wavenumbers 877.77 cm$^{-1}$ and 1372.35 cm$^{-1}$, obtained by the DFT-D method. We have also identified two intense peaks for the simulated Raman spectrum of the 2D single layer $V_N$_d-BN material. One of these peaks corresponds to a vibrational mode ($v_1$) occurring at a wavenumber of 1422.21 cm$^{-1}$, considered as a reference with 100% intensity. The second vibrational mode ($v_2$) appears at 1338.39 cm$^{-1}$ and exhibits approximately 98% of the intensity of the highest Raman active vibrational mode $v_1$. We have also observed other vibrations to the left of this intense peak, but their intensity is less than the $v_1$ vibrational mode. The range of these vibrations is around 50–80% of the intensity of the $v_1$ mode. The Raman active spectrum spread from 100 to 1550 cm$^{-1}$ in this case. However, all other vibrations are comparatively less intense, falling below 50% of the intensity of $v_1$.

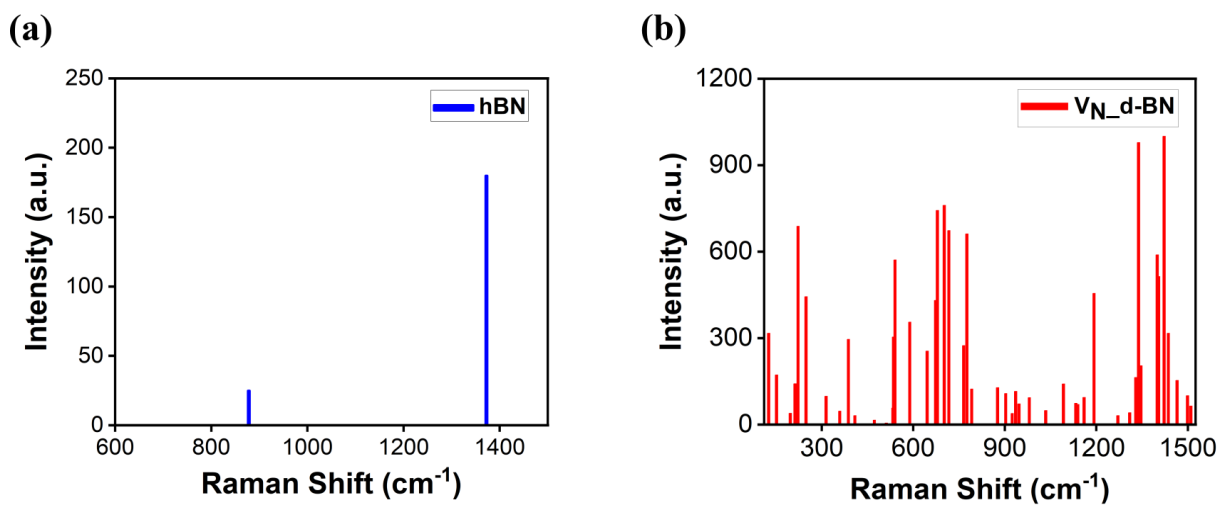

**Fig. 2** Simulated Raman spectra of the (a) pristine 2D monolayer hBN and (b) 2D monolayer $V_N\_d$-BN materials are shown here.

After determining the stable configuration, electronic structure, geometry, and electronic properties of the $V_N\_d$-BN material, we have explored how effective this $V_N\_d$-BN material is as an electrocatalyst for $CO_2RR$. The primary objective of this work is to examine the potential of this $V_N\_d$-BN sheet as an efficient electrocatalytic material for the $CO_2RR$. In this study, we have explored various possible reaction pathways and primary reduction products involved in the $CO_2RR$ utilizing the periodic 2D slab structure of the $V_N\_d$-BN material, which is based on the computational hydrogen electrode (CHE). The complete $CO_2RR$ process includes multiple reaction steps with various intermediates. By studying numerous adsorption sites, we have identified the most stable adsorption structures of the reaction intermediates. In our analysis, we have only focused on the formation of single-carbon intermediates and products such as HCOOH, CO, $CH_3OH$, and $CH_4$. Here, by following an experimental work by Katerina L. et al., we have excluded all other possibilities of the formation of multi-carbon compounds.[40] The main objective of the present theoretical work is to explain the experimental observation performed by Katerina L. et al.[40] During each hydrogenation step, several $CO_2RR$ routes can exist where the H can adsorb on the C or O atom. In this present study, we have discussed the most favorable reaction path of $CO_2RR$ towards various intermediates and products in detail. We have calculated the change in Gibbs free energies for all considered paths, helping us to understand the complete reduction process and determine the optimal path and final achievable products on the surface of the $V_N\_d$-BN electrocatalyst.

The adsorption of linear $CO_2$ molecules is challenging because $CO_2$ is fully oxidized and thermodynamically stable. So, in most of the cases, it cannot be absorbed effectively and spontaneously on the catalytic surface. Therefore, we have considered the adsorption and activation of $CO_2$ molecules on the surface of the catalysts as the first step of $CO_2RR$. Usually, this process requires electron injection into the antibonding $2\pi_u$ orbitals of the $CO_2$ molecule. Therefore, the 2D monolayer $V_N\_d$-BN can capture the $CO_2$ molecule effectively due to its large number of exposed active sites with the improved electron transfer capacity due to free radicals. In the most stable adsorption configuration of the $CO_2$ molecule, one of the oxygen and carbon atoms are bonded with two active sites (Boron), as shown in Fig. 3a. When $CO_2*$ interacts with the $V_N\_d$-BN material at its surface, the C-O bond length significantly increases from 1.160 Å to 1.292 Å obtained by the DFT-D method. After the adsorption of $CO_2*$ onto the $V_N\_d$-BN surface, the equilibrium bond lengths for B-C, B-O, and C-O are found to be 1.591 Å, 1.392 Å, and 1.292 Å, respectively, as reported in Table 1. Now, the linear structure of the $CO_2$ molecule transforms to a V-shape structure with an angle of 120.29˚. This transformation indicates that the $V_N\_d$-BN material effectively activates the $CO_2$ molecule. The value of $\Delta G$ during this reaction step is approximately -0.92 eV, as reported in Table 2, indicating an exothermic nature of the reaction and, hence, thermodynamically favorable. Therefore, the negative value of change in Gibbs free energy of adsorption of $CO_2$* favors the stability of the reaction on the catalytic surface. The optimized lattice parameters, space group symmetry, electronic band gap, and average bond lengths for the $CO_2$ adsorption on the surface of the 2D monolayer $V_N\_d$-BN material under equilibrium conditions are presented in Table 1.

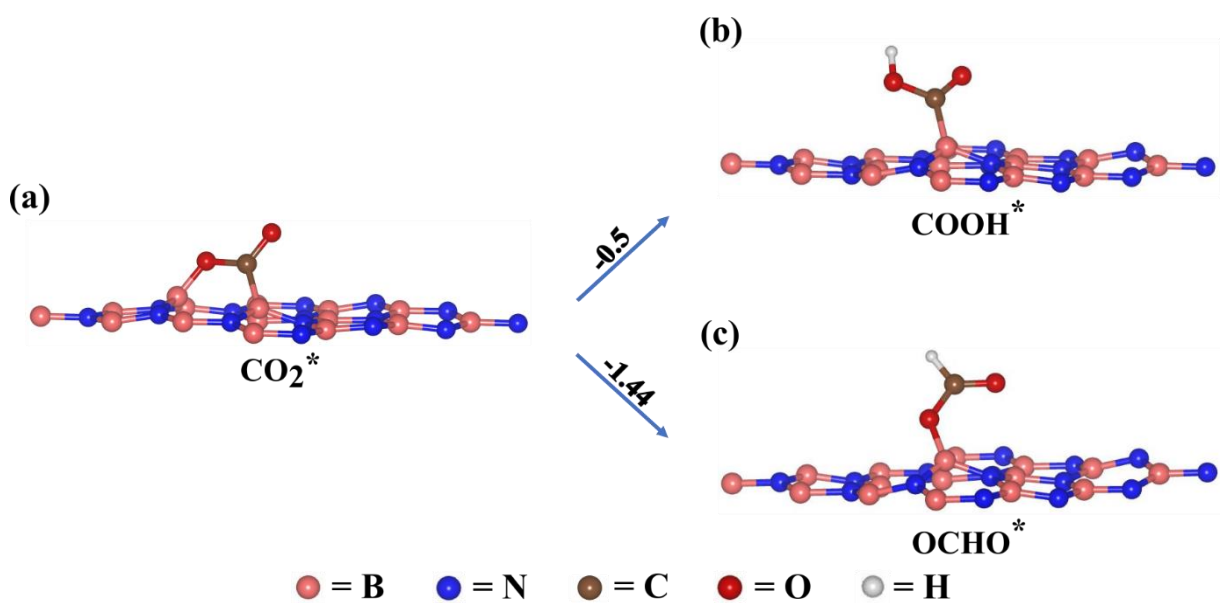

**Fig. 3** Adsorption of $CO_2$ molecule and formation of COOH* and OCHO* intermediates on the surface of $V_N\_d$-BN material with the optimized structures of (a) $CO_2$*, (b) COOH*, and (c) OCHO*.

After the successful adsorption of the $CO_2$ molecule, the first stage of hydrogenation of the adsorbed $CO_2$ molecule can take place either on the C or O site, resulting in the formation of a stable carboxyl group (COOH) or formate group (OCHO) intermediates, respectively. Fig. 3 shows the optimized structures of both the COOH and OCHO intermediates formed during the $CO_2$RR on the surface of $V_N\_dBN$ material. The initial hydrogenation step is crucial in determining the overall reaction path of $CO_2$RR. In this study, we have extensively examined paths of both the COOH and OCHO intermediates for producing formic acid (HCOOH), methanol ($CH_3OH$), and methane ($CH_4$). The potential energy surface diagram (as depicted in Fig. 6) i.e., reaction pathway shows the relative Gibbs free energies for both the considered paths of $CO_2RR$ on the surfaces of the 2D $V_N\_d$-BN material. It also shows the formation of different intermediates and products with C1 species according to the number of hydrogenation steps for both $CO_2RR$ paths.

**CO₂RR through the COOH path**

The formation of COOH* on the surface of $V_N\_d$-BN material occurred through the first hydrogenation step at the O site of the adsorbed V-shaped $CO_2$ molecule as depicted in Figure 3a. Fig. 4 shows the equilibrium structures of all these intermediates of the COOH path with their corresponding values of $\Delta G$. The hydrogenation at the O site does not significantly change the $CO_2$ hydrogenation structure. During this hydrogenation step, the B-O bond splits, and a new O-H bond is formed on the catalytic surface, forming a COOH* intermediate, as shown in Fig. 3b and 4a. In this elementary reaction step, we observed a $\Delta G$ value of -0.5 eV for the COOH* formation on the surface of the 2D monolayer $V_N\_d$-BN material. The negative value of $\Delta G$ indicates that the reaction step is exothermic and spontaneous. In the second hydrogenation step, the hydrogen can bond with a C or O atom of the COOH* intermediate. If the hydrogen bonds with the oxygen atom located distantly from the surface, it produces an $H_2O$ molecule, which leaves the catalytic surface, and only CO* remains on the surface of $V_N\_d$-BN material, as shown in Fig. 4c. In this step, we observed a value of $\Delta G$ around -1.67 eV. The higher negative value of $\Delta G$ indicates that hydrogen can strongly bind to the surface of $V_N\_d$-BN material, making the desorption step difficult. On the other hand, if hydrogen bonds with the carbon atom, it leads to the formation of HCOOH* intermediate on the surface of $V_N\_d$-BN material, as depicted in Fig. 4b. This reaction step shows a positive value of $\Delta G$ around 1.87 eV. This higher positive value of $\Delta G$ signifies a weak hydrogen binding on the surface of $V_N\_d$-BN material, which will create difficulty in adsorption. The negative value of $\Delta G$ for CO* formation confirmed the exothermic nature of the reaction, so it occurs spontaneously on the surface of $V_N\_d$-BN material. In contrast, the positive value of $\Delta G$ for HCOOH* formation indicates the endothermic nature of the reaction, so we require energy for hydrogenation on the catalytic surface. Here, we consider the CO* formation due to its negative value of $\Delta G$ for further hydrogenation. The CO* intermediate plays a crucial role in the $CO_2RR$ as its ability to adsorb on a catalyst strongly influences whether CO becomes the primary product or undergoes additional hydrogenation. If the value of ΔG for forming the CO* is small, it will desorb as a CO molecule from the catalytic surface as the final product. However, if the value of ΔG for forming the CO* on the surface of $V_N\_d$-BN material is moderate, it will more favorably generate final products containing more than two electrons, such as CH$_3$OH and CH$_4$. Here, the value of ΔG is around -1.67 eV, indicating that CO* hydrogenates in further reaction steps to produce CH$_3$OH or CH$_4$ as a final product.

In the third hydrogenation step, the adsorbed CO* molecule undergoes further hydrogenation to yield COH* and CHO* rather than undergoing desorption. On the $V_N\_d$-BN surface, the value of ΔG required for CO* hydrogenation to form CHO* intermediate is 1.98 eV, as shown in Fig. 4d. In contrast, the value of ΔG required for CO* hydrogenation to form COH* intermediate is 4.32 eV, as shown in Fig. 4e. The positive value of ΔG for both cases indicate the endothermic nature of the reaction, so we require energy for further hydrogenate the CO* intermediate on the surface of the 2D single layer $V_N\_d$-BN material. This higher positive value of ΔG signifies a weak hydrogen binding on the surface of $V_N\_d$-BN material, which will create difficulty in adsorption. Therefore, the higher positive value of ΔG hinders the further hydrogenation of CO* intermediate. It leads to a thermodynamically unfavorable reaction as a higher applied potential will be required to generate CHO* and COH* intermediate along the COOH path.

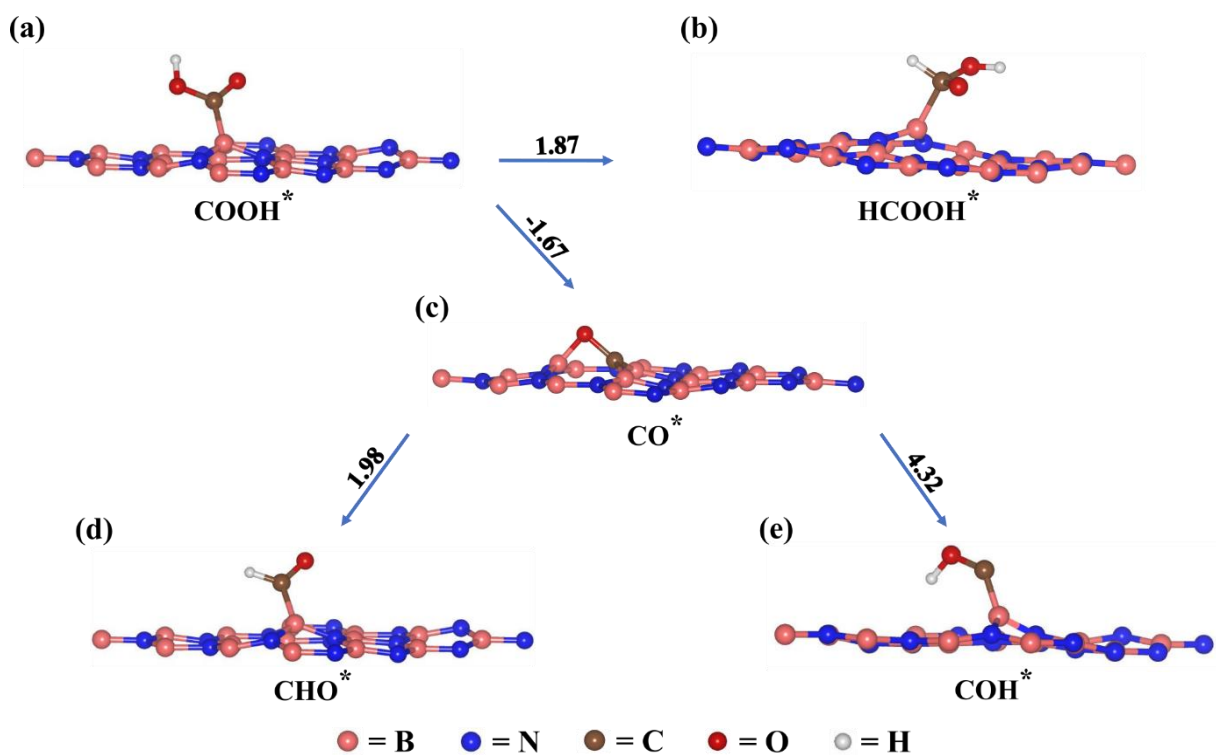

**Fig. 4** Formation of HCOOH*, CO*, CHO*, and COH* intermediates on the surface of the 2D monolayer $V_{N\_}$d-BN material with the equilibrium structures of (a) COOH*, (b) HCOOH*, (c) CO*, (d) CHO*, and (e) COH*.

## CO₂RR through the OCHO path

Now, we initiate the CO₂RR using an activated CO₂ molecule on the surface of the $V_{N\_}$d-BN material for the OCHO path. Fig. 5 shows the equilibrium structures and values of $\Delta G$ for possible intermediates of the OCHO path during the CO₂RR process. The first hydrogenation step occurs at the C site of adsorbed CO₂ molecules and splits the C-B bond. Then hydrogen and the carbon atom form a C-H bond, resulting in the formation of an OCHO* intermediate, as shown in Fig. 5b. In this hydrogenation step, we observed a value of $\Delta G$ around -1.44 eV for OCHO* intermediate formation, as reported in Table 2. The negative values of $\Delta G$ indicate that the reaction step is exothermic and spontaneous. It is important to note here that the value of $\Delta G$ for OCHO* formation ($\Delta G = -1.44$ eV) is higher than the value of $\Delta G$ for COOH* formation ($\Delta G = -0.50$ eV), so the formation of OCHO* intermediate is considered thermodynamically more favorable. Also, the higher energy released during OCHO* formation indicates the effective activation of the CO₂ molecule on the surface of the 2D monolayer $V_{N\_}$d-BN material. Therefore, we considered the formation of OCHO* intermediate on the surface of $V_{N\_}$d-BN material as a primary product for further reduction. The second hydrogenation step occurs at the distantly located O site on the surface of the OCHO* intermediate. This hydrogenation step results in the formation of an OCHOH* intermediate on the catalytic surface, as shown in Fig. 5c. The calculated value of $\Delta G$ for this reaction step is 1.80 eV, as reported in Table 2. The positive values of $\Delta G$ indicate the endothermic nature of the reaction, so we require energy to hydrogenate the OCHOH* intermediate further on the surface of the $V_{N\_}$d-BN material. So, this positive change in the value of $\Delta G$ provides some resistance to further hydrogenation on the catalytic surface. Still, this path is more favorable than the previously discussed COOH path. Therefore, we consider the formation of an OCHOH* intermediate for further reduction on the surface of the $V_{N\_}$d-BN material.

In the third hydrogenation step, the hydrogen combines with the distantly located oxygen atom of the O-H site in the OCHOH* intermediate, and the H₂O molecule is desorbed from the catalytic surface. Here, only CHO* remains on the surface of $V_{N\_}$d-BN material, as shown in Fig. 5d. The calculated value for $\Delta G$ is -0.55 eV for CHO* intermediate formation, as reported in Table 2. The negative value of ΔG indicates the exothermic nature of this reaction step. So, it spontaneously occurs on the catalytic surface. It is important to note that the OCHO path is more favorable than the COOH path here, but at the end of both reaction paths, we get CHO* as a common reaction intermediate. In the fourth hydrogenation step, the hydrogenation of CHO* can lead to the formation of CHOH* and $OCH_2$* intermediates. Now, if the hydrogen atom bonded to the C site on the surface of the CHO*, it forms an $OCH_2$* intermediate. But if a hydrogen atom bonded to the O site on the surface of the CHO* intermediate, it forms a CHOH intermediate. The value of ΔG for the CHOH* intermediate formation is calculated to be 1.21 eV, while the value of ΔG for the $OCH_2$* intermediate formation is around 0.11 eV obtained by the DFT-D method. Here, the change in Gibbs free energies for both the reaction steps is positive, which shows the endothermic nature of the reaction steps of these intermediate states. However, the value of ΔG for the $OCH_2$* formation is much less than that for the HCOH* formation, so we consider the formation of the $OCH_2$* intermediate on the surface of $V_N$_d-BN for further reduction. The equilibrium structure of $OCH_2$* intermediate on the surface of the 2D monolayer $V_N$_d-BN material is depicted in Fig. 5e.

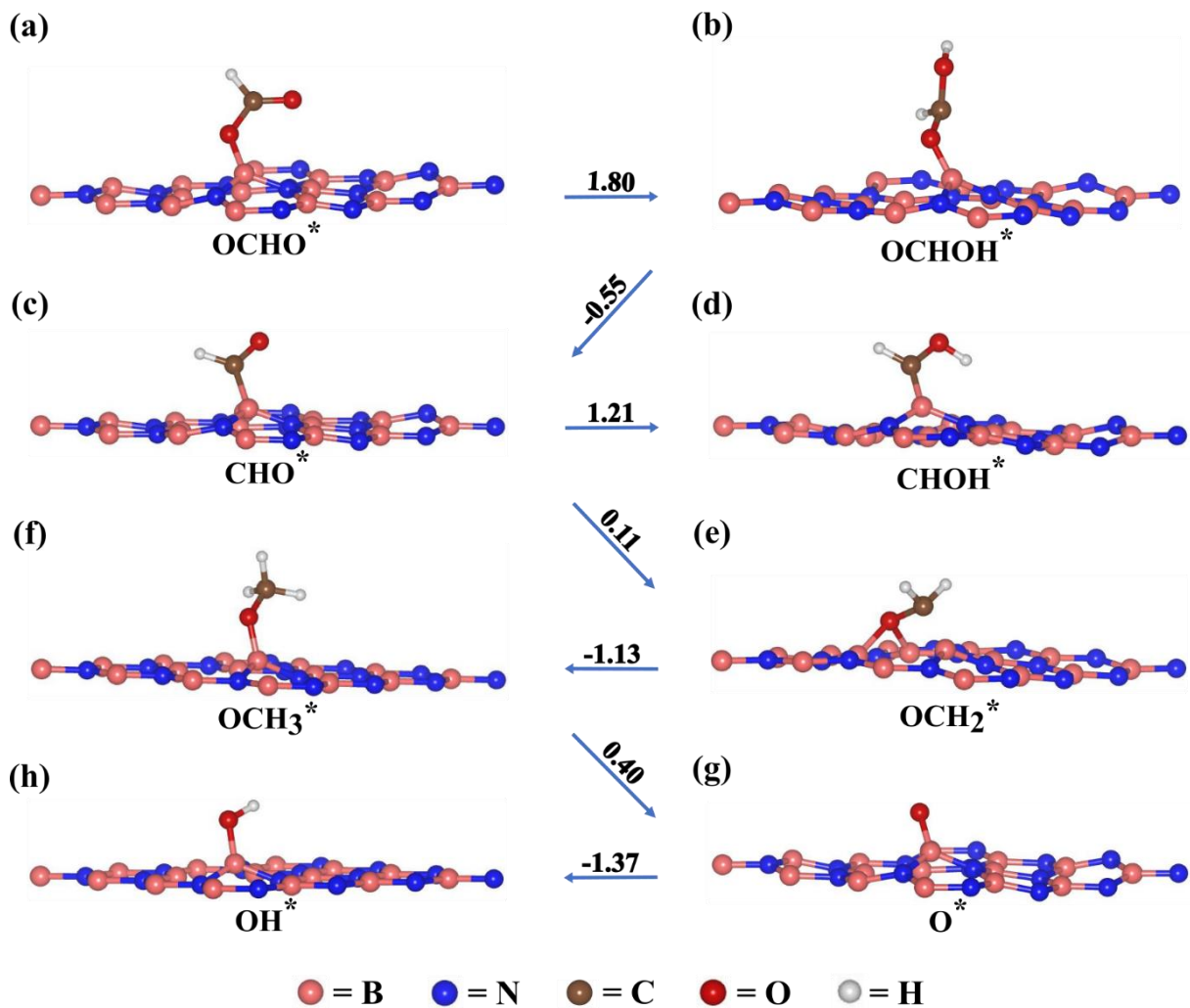

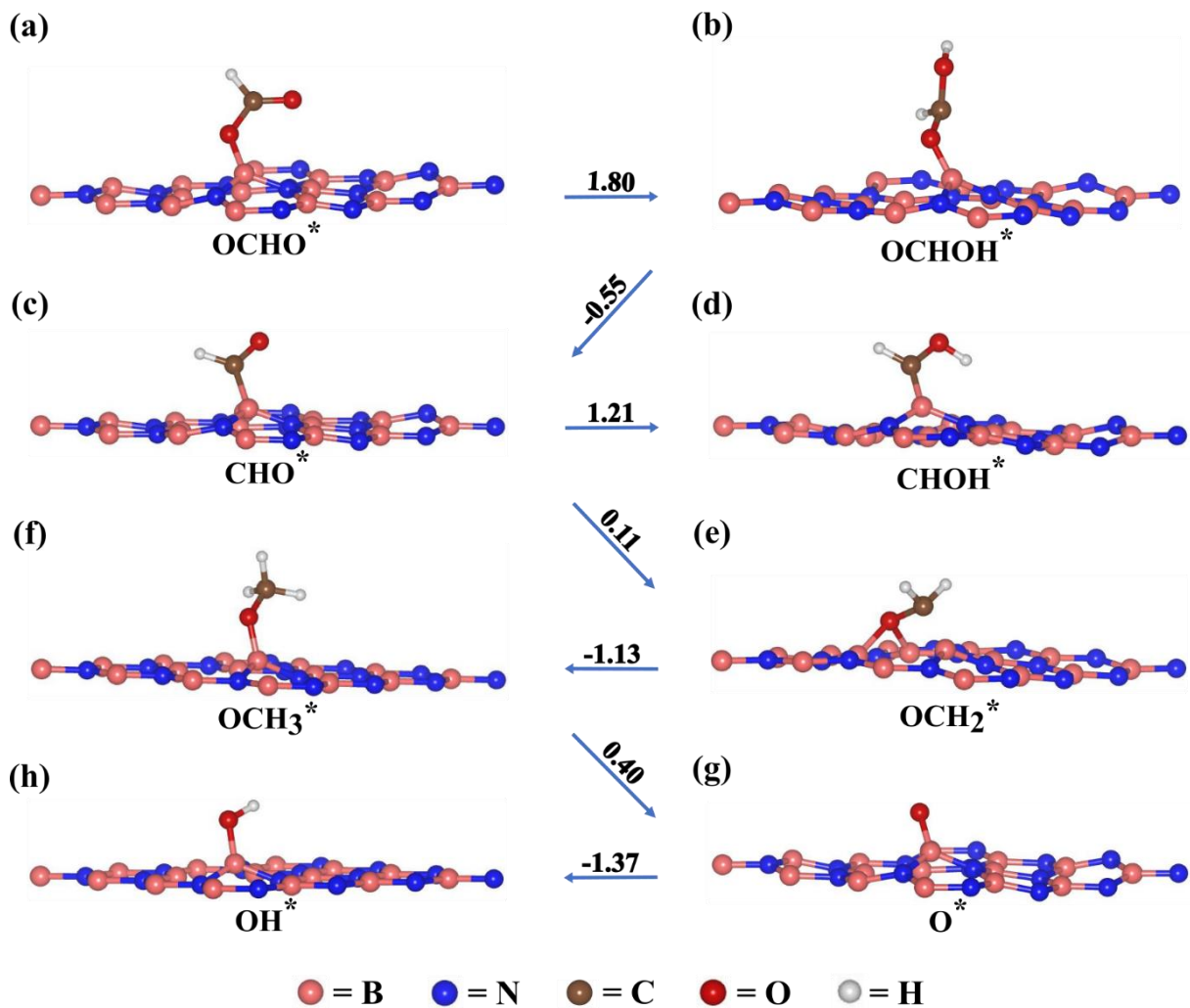

🔴 = B    🔵 = N    🟤 = C    🔴 = O    ⚪ = H

**Fig. 5** Adsorption of $CO_2$ molecule and formation of various intermediates on the surface of 2D monolayer $V_N$_d-BN material with the equilibrium structures of (a) $CO_2$*, (b) OCHO*, (c) OCHOH*, (d) CHO*, (e) $OCH_2$*, (f) $OCH_3$*, (g) O*, and (h) OH*.

In the fifth hydrogenation step, the intermediate $OCH_2$* undergoes further hydrogenation, where a hydrogen atom is adsorbed at the C site to form the $OCH_3$* reaction intermediate. Fig. 5f shows the equilibrium structure of the $OCH_3$* intermediate on the surface of the 2D monolayer $V_N$_d-BN material. The calculated value of $\Delta G$ for the $OCH_3$* intermediate formation on the catalytic surface is about -1.13 eV, as reported in Table 2. The negative value of $\Delta G$ indicates the exothermic nature of the reaction, which spontaneously occurs on the catalytic surface. In the sixth step, the $OCH_3$* intermediate can undergo hydrogenation on the surface of the $V_N$_d-BN material with two possibilities. The first possibility is where hydrogen gets bonded at the O site of the $OCH_3$* intermediate, which leads to the formation of a $CH_3OH$ molecule. This $CH_3OH$ molecule gets desorbed from the surface of the $V_N$_d-BN material as a final product. The second possibility is the desorption of a $CH_4$ molecule from the catalytic surface. Now, only a single oxygen atom remains to produce an $H_2O$ molecule as a final product after two more successive hydrogenation steps. The hydrogenation of the $OCH_3*$ intermediate to form $CH_3OH*$ is accompanied by a value of $\Delta G$ around 1.37 eV. In comparison, the formation of the O* intermediate with desorption of the $CH_4$ molecule has a change in Gibbs free energy around 0.40 eV, as reported in Table 2. The value of $\Delta G$ for both steps is positive, which shows the endothermic nature of both reaction steps. The formation of an O* intermediate is more favorable than the desorption of a $CH_3OH$ molecule due to a comparatively smaller value of $\Delta G$. Hence, we consider an O* intermediate with the desorption of a $CH_4$ molecule on the catalytic surface for further reduction. The optimized structure of O* intermediate on the surface of the 2D monolayer $V_N\_d$-BN material is shown in Fig. 5g.

The intermediate O* undergoes further hydrogenation in the seventh step, where hydrogen bonds with the remaining oxygen on the catalytic surface to form the OH* intermediate, as shown in Fig. 5h. The value of $\Delta G$ for this reaction step is around -1.37 eV, as reported in table 2. The negative value of $\Delta G$ for OH* formation confirmed the exothermic nature of the reaction, so it occurs spontaneously on the catalytic surface. The remaining OH* intermediate bonded with hydrogen in the last hydrogenation step, forming an $H_2O$ molecule that desorbed from the catalytic surface for the next cycle. The value of $\Delta G$ for this hydrogenation step is around 2.18 eV, as reported in Table 2. So, the hydrogenation of OH* is the highly endothermic reaction step and is also considered a rate-limiting step of $CO_2RR$. The formation of $H_2O$ from the OH* intermediate serves as a crucial step that contributes to the overall reaction and plays a pivotal role in facilitating subsequent cycles of the reaction. The value of change in Gibbs free energy for $H_2O$ formation on the catalytic surface is highly positive, but still, it is a crucial reaction step because it provides an active catalytic surface for the next cycle of $CO_2RR$. The formation of $H_2O$ at the end of the reaction cycle restores the active sites on the catalytic surface, preparing it for initiating further $CO_2RR$ cycles. The formation of $H_2O$ may create a thermodynamic barrier, but it is a crucial intermediate in restoring the catalyst surface. Therefore, it draws our attention to its importance in sustaining and enhancing the efficiency of $CO_2RR$ processes.

The potential energy surface (PES) diagram illustrates the energetics of the $CO_2RR$ mechanism on the surface of $V_N\_d$-BN material. We present the two prominent paths in the PES diagram by studying all possible routes. The first is the COOH* path, represented by red color, and the second is the OCHO* path, represented by blue color in the PES diagram, as shown in Fig. 6. Both pathways represent a series of intermediate steps leading to the final product. We can easily understand which path is more thermodynamically favorable by analyzing the relative Gibbs free energies along these $CO_2RR$ pathways. The PES diagram shows which path is more accessible and gives essential information about how likely it will happen. This analysis showed us that the OCHO* path exhibits more favorable than the COOH* path. This conclusion results from observing the change in Gibbs free energy for the formation of each reaction intermediate along both paths. Consequently, the OCHO* path emerges as the preferred route for the $CO_2RR$ process on the surface of the 2D monolayer $V_{N}$_d-BN material. Various factors, such as the specific electronic and structural properties of the 2D monolayer $V_{N}$_d-BN material, favorably influence the adsorption and activation of $CO_2$ molecules and intermediates along the OCHO* path. Thoroughly analyzing the PES diagram with the react pathways offers valuable insights into how the $CO_2RR$ process happens on the surface of $V_{N}$_d-BN material. It highlights the importance of the OCHO* path in reacting toward favorable results which is accord with the experimental observation performed by Katerina et al.[40]

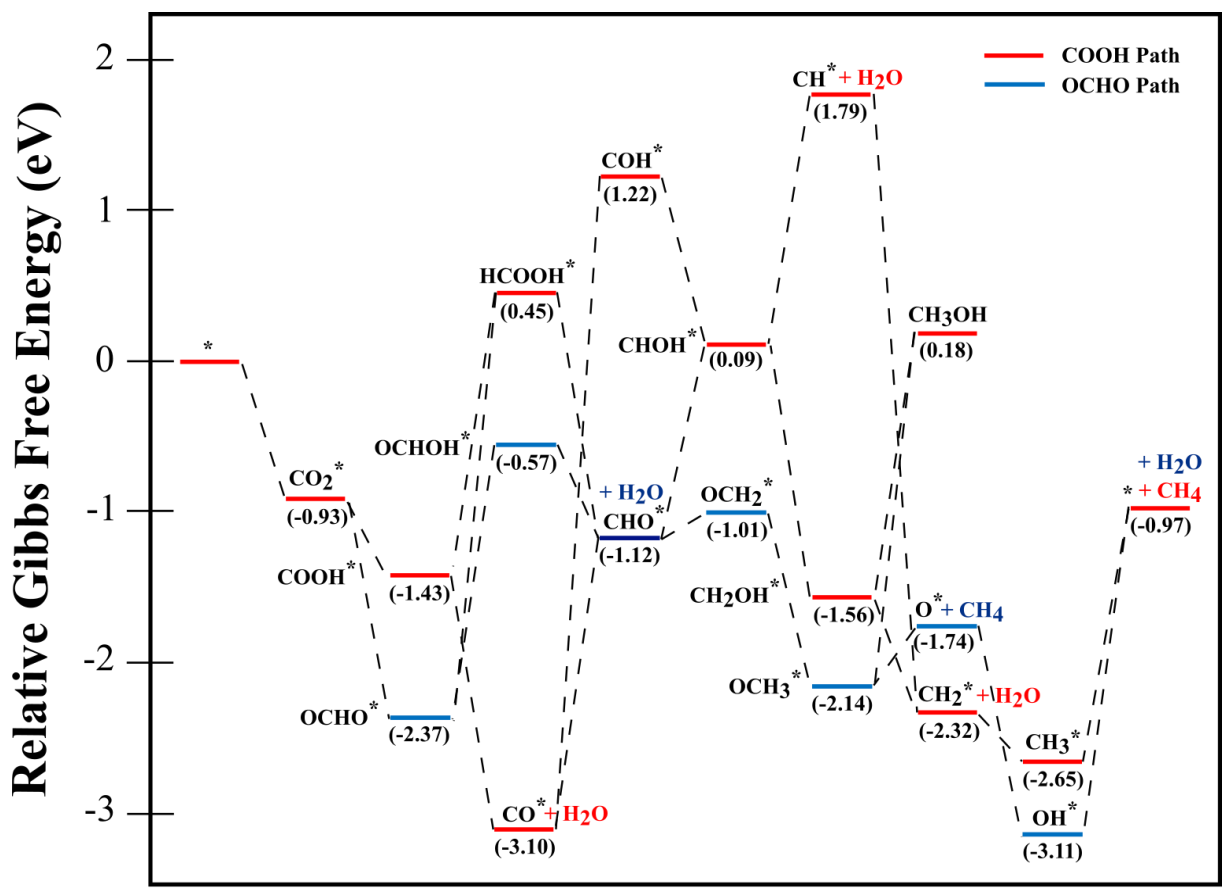

**Fig. 6** Relative Gibbs free energy diagram for the CO$_2$RR on the surface of the 2D monolayer V$_N$_d-BN material.

**Table 1** Equilibrium structural parameters, lattice constants, and electronic band gap (E$_g$) parameters of various systems of the CO$_2$RR.

| Reaction intermediate | Lattice parameters (Å) | Interfacial angles in degree | Space group & symmetry | Electronic band gap (E$_g$ in eV) | Average bond distance (in Å) | | | | |
|---|---|---|---|---|---|---|---|---|---|
| | | | | | B-C | B-O | C-O | C-H | O-H |
| CO$_2$* | a = 7.509, b = 7.540 | $\alpha = \beta = 90$ $\gamma = 120.480$ | *P1* | 0 | 1.591 | 1.392 | 1.292 | - | - |
| OCHO* | a = 7.215, b = 7.472 | $\alpha = \beta = 90$ $\gamma = 118.882$ | *P1* | 5.30 | - | 1.417 | 1.281 | 1.096 | - |
| OCHOH* | a = 7.209, b = 7.476 | $\alpha = \beta = 90$ $\gamma = 118.879$ | *P1* | 0 | - | 1.382 | 1.374 | 1.091 | 0.971 |
| CHO* | a = 7.259, b = 7.474 | $\alpha = \beta = 90$ $\gamma = 119.041$ | *P1* | 0 | 1.604 | - | 1.228 | 1.113 | - |
| OCH$_2$* | a = 7.361, | $\alpha = \beta = 90$ | *P1* | 0 | 1.576 | 1.559 | 1.470 | 1.089 | - |

| | b = 7.549 | γ = 119.192 | | | | | | | |
|---|---|---|---|---|---|---|---|---|---|
| OCH$_3$* | a = 7.497, b = 7.497 | α = β = 90 γ = 120.000 | *P1* | 0 | - | 1.390 | 1.417 | 1.092 | - |
| O* | a = 7.209, b = 7.483 | α = β = 90 γ = 118.796 | *P1* | 1.68 | - | 1.360 | - | - | - |
| OH* | a = 7.497, b = 7.497 | α = β = 90 γ = 120.000 | *P1* | 0 | - | 1.396 | - | - | 0.968 |

**Table 2** Change in Gibbs free energy (ΔG in eV) and relative Gibbs free energy of all the intermediates during the CO$_2$RR performed on the surface of the 2D monolayer V$_N$_d-BN material is reported here.

| Various CO$_2$RR Steps | ΔG (eV) | Relative free energy (eV) |
|---|---|---|
| V$_N$_d-BN*   →   CO$_2$* | -0.93 | -0.93 |
| CO$_2$*   →   OCHO* | -1.44 | -2.37 |
| OCHO*   →   OCHOH* | 1.80 | -0.57 |
| OCHOH*   →   CHO* + H$_2$O | -0.55 | -1.12 |
| CHO*   →   OCH$_2$* | 0.11 | -1.01 |
| OCH$_2$*   →   OCH$_3$* | -1.13 | -2.14 |
| OCH$_3$*   →   O* + CH$_4$ | 0.40 | -1.74 |
| O*   →   OH* | -1.37 | -3.11 |
| OH*   →   V$_N$_d-BN* + H$_2$O | 2.14 | -0.97 |

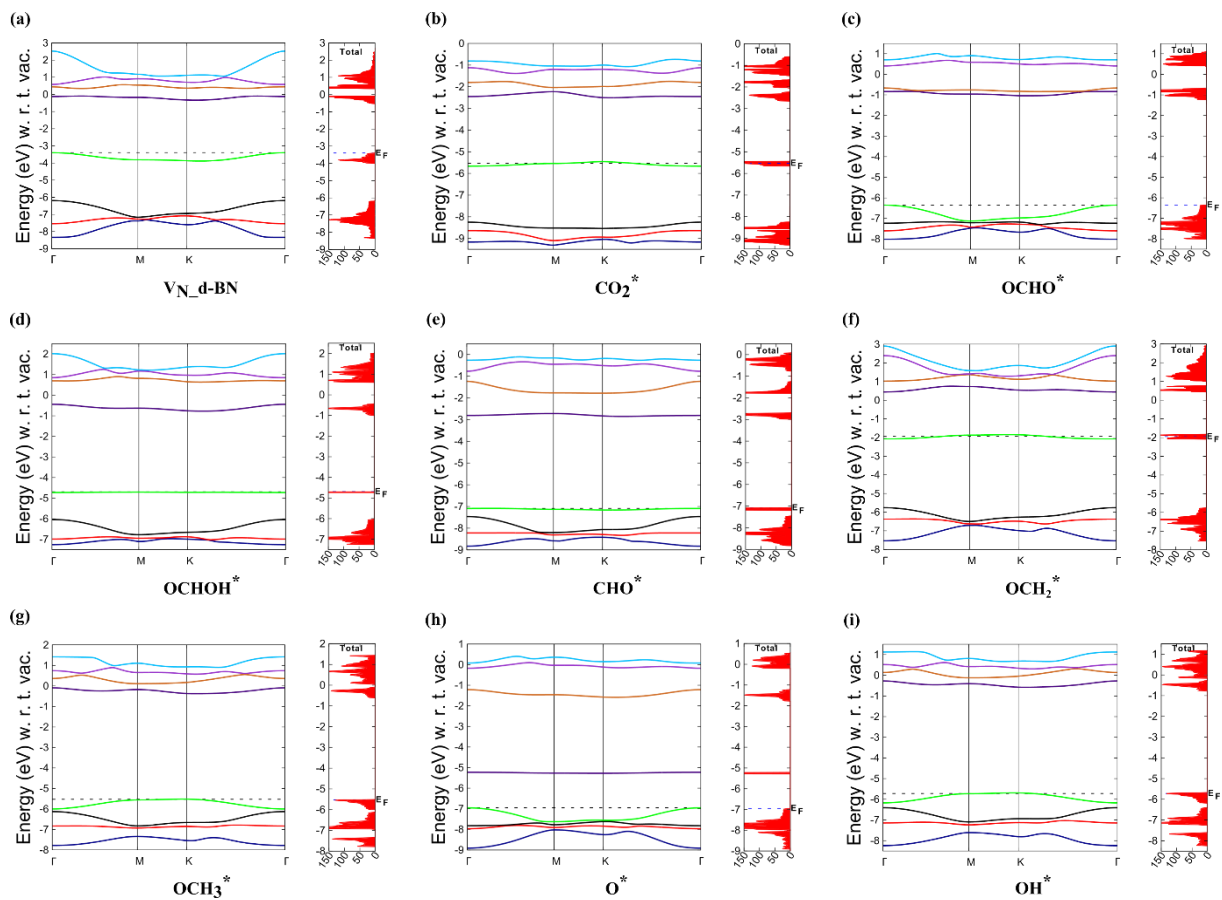

**Fig. 7** Schematic illustration of the electronic band structures and total density of states of the 2D monolayer $V_N\_d$-BN material and all the reaction intermediates along the OCHO path during $CO_2RR$ process are plotted here: (a) $V_N\_d$-BN, (b) $CO_2^*$, (c) OCHO*, (d) OCHOH*, (e) CHO*, (f) $OCH_2^*$, (g) $OCH_3^*$, (h) O*, and (i) OH*

In our investigation of the $CO_2RR$ mechanism on the surface of the 2D monolayer $V_N\_dBN$ material, we employed the DFT-D method to explore the electronic properties to understand the catalytic activity of the material. Specifically, we focused on the band structure, electronic band gap ($E_g$), Fermi energy ($E_F$) level, and total density of states (DOS) for the $V_N\_d$-BN material and each reaction intermediate along the OCHO path during $CO_2RR$. The band structure analysis helps us understand the electronic behavior of a material, highlighting its potential for catalytic activity. Meanwhile, the DOS calculations provide detailed information on the density of electronic states, helping us to observe the availability and distribution of electronic states. To precisely represent the electronic structure, we selected the k-vector path along the highly symmetric directions, named ***Γ−M−M−Γ***, within the first Brillouin zone for band structure plots. Using the band structure calculations, we have plotted eight electronic energy bands. Four out of these eight energy bands lying above the Fermi energy ($E_F$) level are named conduction bands, whereas the remaining four lying below the Fermi energy level are named valence bands. The Fermi energy ($E_F$) level is represented by the dotted blue line in the computed band structures and DOS calculations as depicted in Figure 7a-I for all the reaction intermediates involved in the subject reaction along with the 2D monolayer $V_N\_$dBN material. We have computed the value of Fermi energy level ($E_F$) at -3.39 eV in the electronic band structures and total DOS of the 2D monolayer $V_N\_$d-BN material, as shown in Fig. 7a. We observed a tiny fraction of electron density in the total DOS calculations just below the Fermi energy level.

In the electronic band structure calculations, during the initial hydrogenation step, where the $CO_2$ molecule adsorbs on the surface of the $V_N\_$d-BN material, the Fermi energy ($E_F$) level shifts toward the conduction band. Eventually, one of the electronic energy bands belonging to the conduction bands crosses the Fermi energy ($E_F$) level, indicating a conductive nature of $CO_2$*. The DOS calculations reveal a substantial electron density of states around the Fermi level ($E_F$), which also confirms the conductive nature of $CO_2$* reaction intermediate formed during the $CO_2RR$, as shown in Fig. 7b. This characteristic promotes electron propagation during the reaction, and it helps to improve the efficiency of the $CO_2RR$ mechanism. This conductive nature is observed consistently in the further hydrogenation steps involving OCHOH*, CHO*, $OCH_2$*, $OCH_3$*, and OH* on the surface of the 2D monolayer $V_N\_$d-BN material, as shown in Fig. 7d, 7e, 7f, 7g, and 7i, respectively. We observe similar trends in the electronic band structure calculations of these intermediate states, where one of the electronic energy bands of the valence bands shifts towards the Fermi energy level and crosses it, confirming the conductive nature of these intermediates. We also calculated the total DOS for these intermediates and observed the electron density around the Fermi energy level, which also indicates the conductive nature of these intermediates. We also computed the electronic band structure and total DOS of the remaining intermediates formed during the $CO_2RR$. The electronic band structure calculations of the reaction intermediates OCHO* and O* reaction steps reveal the presence of an energy bandgap in these systems, as shown in Fig. 7c and 7h, respectively. In the DOS calculations, we observed a small fraction of electron density just below the Fermi energy level for these systems. Fig. 7a-i shows the electronic band structure and total DOS of the 2D monolayer $V_N\_$d-BN material along all the reaction intermediates involved in the subject reaction. Fig. 6 illustrates the PES diagram, in which we present the relative Gibbs free energies of the two prominent reaction pathways (COOH and OCHO) of the $CO_2RR$ mechanism on the surface of the 2D monolayer $V_N\_$d-BN material. Table 1 provides a detailed comparison of various reaction intermediates throughout the CO₂RR process, including their equilibrium lattice constants, space group symmetry, shift of Fermi energy level, electronic band gap, and average bond lengths. Table 2 presents the change in Gibbs free energy (ΔG) and relative Gibbs free energy for each intermediate involved in the CO₂RR process performed on the surface of the 2D single layer $V_N$_d-BN material.

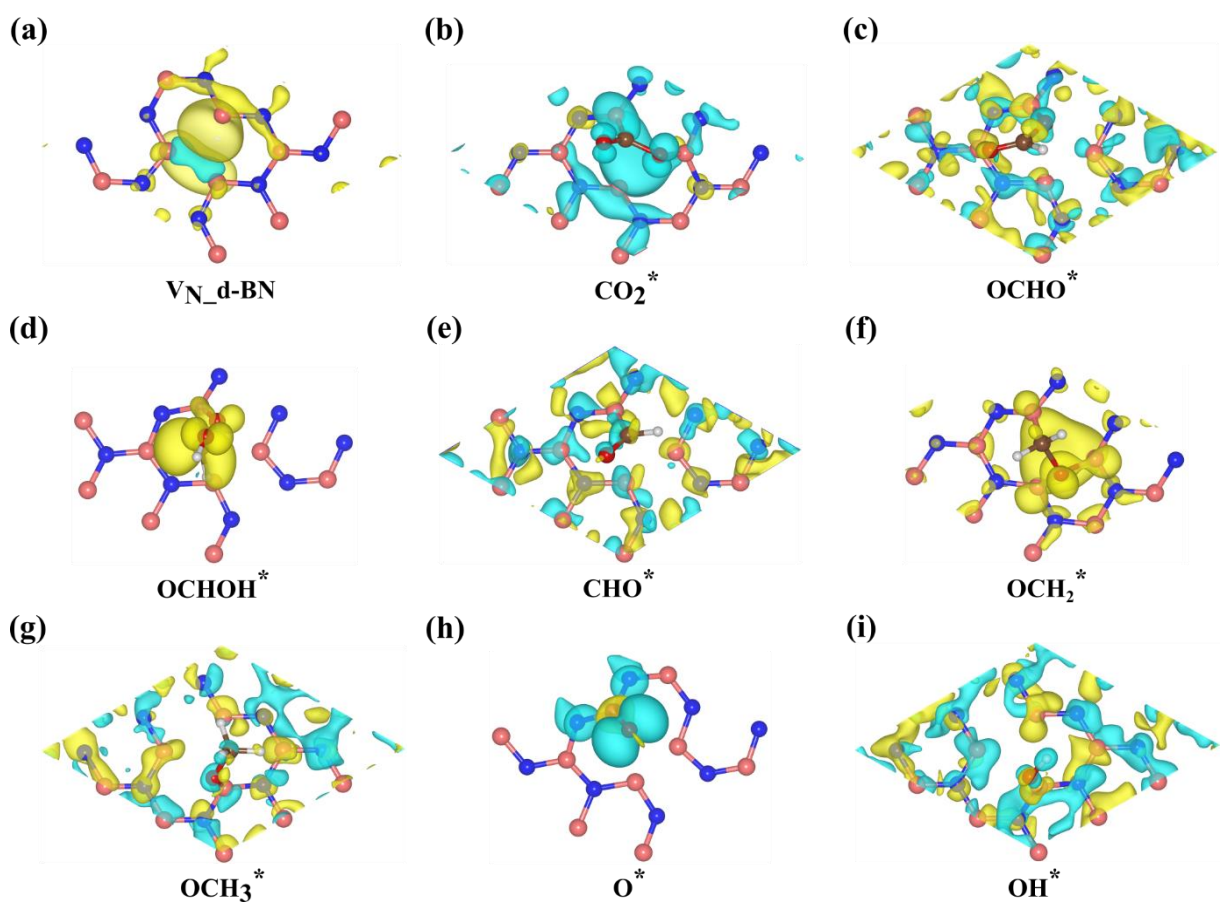

**Fig. 8** Electron spin density calculations of $V_N$_d-BN material and all intermediates formed along the OCHO path during the CO₂RR process: (a) $V_N$_d-BN, (b) CO₂*, (c) OCHO*, (d) OCHOH*, (e) CHO*, (f) OCH₂*, (g) OCH₃*, (h) O* and (i) OH*.

In this study, we also calculated the electron spin densities of the 2D single layer $V_N$_d-BN material along all the intermediate states formed during the CO₂RR process by using the same DFT-D method, as depicted in Fig. 8. The electron spin densities offer crucial insights into the electronic structure and reactivity of these systems, which are essential for understanding their catalytic activity towards the CO₂RR. The significance of these calculations lies in their ability to elucidate the distribution of unpaired electrons, which directly influences the reactivity of materials. The distribution of electron spin density is crucial to show how the spin polarization is within the molecular orbitals. The analysis of spin densities unveils distinct patterns of electron localization and delocalization, highlighting the pivotal role of adsorption sites on the surface of $V_N\_d$-BN material in modulating electron transfer kinetics. The spin density represents the mismatch between the numbers of spin-up ($\alpha$) and spin-down ($\beta$) electrons employed to estimate the presence of unpaired electrons. Under these conditions of unequal distribution of spin-up and spin-down electrons, spin polarization comes into the picture, which also results in non-zero spin density. The spin density functional theory (SDFT) extends DFT to include magnetic fields alongside scalar external potentials from the nuclei. Two space functions, spin density (s) and electron density ($\rho$), serve as fundamental variables in SDFT. The spin density reflects interactions among electron spins within a system and plays a crucial role in understanding the magnetic phenomena of the material.[56] In Fig. 8, the $\alpha$-spin electrons, i.e., the positive component of the electronic wave function, are represented by the highlighted yellow color, while the $\beta$-spin electrons, i.e., the negative component of the electronic wave function, are represented by the sky-blue color. The spin density distribution within a molecule is crucial for understanding the propagation of spin polarization in molecular complexes and crystals. It is helpful to understand various magnetic interactions, which are a function of molecular orientation as well as packing. Hence, we can say that electron density plays a crucial role in electrocatalysis during the $CO_2RR$ to facilitate electron transfer in the reaction mechanism.

## Conclusions

In our current work, we have investigated the electrocatalytic activities of the defective 2D monolayer hexagonal boron nitride (d-BN) towards $CO_2RR$ which is consistent with the experimental observation. By introducing a single nitrogen-vacancy ($V_N$) defect in the pristine 2D monolayer hBN, we have examined the electrocatalytic activity of the 2D single layer $V_N\_d$-BN material towards the $CO_2RR$. We have simulated the Raman spectroscopy of the $V_N\_d$-BN material to determine its overall stability to perform the $CO_2RR$. We have followed the DFT-D method to computationally explore the structural and electronic properties of the 2D monolayer $V_N\_d$-BN material as well as for all the intermediate states formed during the $CO_2RR$. We have calculated the electronic properties of the $V_N\_d$-BN material towards all the possible reaction intermediates formed during the $CO_2RR$, including its band structure and total DOS at the equilibrium position. The findings of our current work suggest that after the introduction of the $V_N$ defect in the pristine 2D monolayer hBN, the electronic band gap of the 2D monolayer $V_N\_d$-BN material is reduced to 3.0 eV. After determining the stable configuration and electronic properties of the $V_N\_d$-BN material, we have explored various possible reaction pathways and primary reduction products involved in the $CO_2RR$. In the present investigation, we have considered a periodic structural slab structure of the $V_N\_d$-BN material and $CO_2RR$ process has been investigated based on the computational hydrogen electrode (CHE). We have also calculated the electron spin density of all the systems which helps us to locate the unpaired electrons around the defective reason of the $V_N\_d$-BN material. This DFT-D approach was consistently applied throughout the calculations to the precise depth of Gibbs free energy during the $CO_2RR$ process. We have explored the two prominent reaction pathways such as COOH* and OCHO* on the surface of the 2D single layer $V_N\_d$-BN material. Both the reaction pathways represent a series of intermediate steps leading to the final product. The results of our analysis showed that the OCHO* path exhibits more favorable than the COOH* path. Also, we found that the 2d monolayer $V_N\_d$-BN material is more likely to generate $CH_4$ as the final product rather than CO, HCOOH, and $CH_3OH$ during the $CO_2RR$ process. The hydrogenation of OH* intermediate is the rate-limiting step. Overall, the $V_N\_d$-BN material demonstrates significant activity and selectivity as an electrocatalyst for the reduction of $CO_2$ to $CH_4$. Hence, the 2D monolayer $V_N\_d$-BN exhibits promising electrocatalytic activity for $CO_2RR$ with substantially improved reaction kinetics.

## Conflicts of Interest:

The authors have no additional conflicts of interest.

## AUTHOR INFORMATION

**Corresponding Author**
**Dr. Srimanta Pakhira** − *Theoretical Condensed Matter Physics and Advanced Computational Materials Science Laboratory, Department of Physics, Indian Institute of Technology Indore (IIT Indore), Simrol, Khandwa Road, Indore, Madhya Pradesh 453552, India.*

*Theoretical Condensed Matter Physics and Advanced Computational Materials Science Laboratory, Centre of Advanced Electronics (CAE), Indian Institute of Technology Indore, Indore, MP 453552, India.*
ORCID: orcid.org/0000-0002-2488-300X.
Email: spakhira@iiti.ac.in or spakhirafsu@gmail.com

**Authors**

**Mr. Lokesh Yadav** − *Theoretical Condensed Matter Physics and Advanced Computational Materials Science Laboratory, Department of Physics, Indian Institute of Technology Indore (IIT Indore), Simrol, Khandwa Road, Indore, Madhya Pradesh 453552, India.*

## Acknowledgement:

This work was financially supported by the Science and Engineering Research Board-Department of Science and Technology (SERB-DST), Government of India, under Grant No. CRG/2021/000572. We thank the CSIR, Govt of India for providing the research funds under the scheme no. 22/0883/23/EMR-II. Dr Srimanta Pakhira acknowledges the SERB-DST, Government of India, for providing his Early Career Research Award (ECRA) under project number ECR/2018/000255 and for providing the computer cluster. Dr Pakhira also recognizes the SERB-DST for providing the highly prestigious Ramanujan Faculty Fellowship under scheme number SB/S2/RJN-067/2017 and providing the highly prestigious Core Research Grant (CRG), SERB-DST, Govt. of India under the scheme number CRG/2021/000572. Ms. Lokesh thanks the CSIR, Govt. of India, and Govt. of India for providing his doctoral fellowship under scheme no. CSIRAWARD/JRF-NET2022/11898. The author would like to acknowledge the SERB-DST for providing computing clusters and programs.

## Author Contributions:

Dr. Pakhira designed the project, and he conceived the complete idea of this current research project work, Mr. Lokesh Yadav computationally studied the electronic structures and properties of the 2D monolayer hBN and $V_N$_d-BN materials. Dr. Pakhira and Mr. Lokesh Yadav explored the whole reaction paths, transition states, and reaction barriers. They explained the $CO_2RR$ mechanism by the DFT Quantum Mechanical calculations. Dr. Pakhira and Mr. Lokesh wrote the whole manuscript and prepared all the tables and figures in the manuscript. Dr. Pakhira and Mr. Lokesh interpreted and analyzed the computed results, and Dr. Pakhira supervised the project work.

15215–15232.